\documentstyle[pre,aps,eqsecnum,multicol,epsf]{revtex}

\def\beginmcols
{\begin{multicols}{2}%
\narrowtext%
}
\def\endmcols{%
\end{multicols}%
\widetext%
}

\begin{document}

\title{\Large\bf Elastically coupled molecular motors}
\author{Andrej Vilfan, Erwin Frey and Franz Schwabl}
\address{
Institut f\"ur Theoretische Physik,
Physik-Department der Technischen Universit\"at M\"unchen, \\
James-Franck-Stra\ss e, D-85747 Garching, Germany}
\date{\today}
\maketitle


\begin{abstract}
  We study the influence of filament elasticity on the motion of collective
  molecular motors. It is found that for a backbone flexibility exceeding a
  characteristic value (motor stiffness divided through the mean displacement
  between attached motors), the ability of motors to produce force reduces as
  compared to rigidly coupled motors, while the maximum velocity remains
  unchanged. The force-velocity-relation in two different analytic
  approximations is calculated and compared with Monte-Carlo simulations.
  Finally, we extend our model by introducing motors with a strain-dependent
  detachment rate. A remarkable crossover from the nearly hyperbolic shape of
  the Hill curve for stiff backbones to a linear force-velocity relation for
  very elastic backbones is found. With realistic model parameters we show that
  the backbone flexibility plays no role under physiological conditions in
  muscles, but it should be observable in certain {\it in vitro} assays.
\end{abstract}


\beginmcols

\section{Introduction}
Molecular motors play a key role in a variety of biological processes like
muscle contraction, intracellular transport, cell locomotion, flagellar
rotation etc.\ \cite{alberts}. Despite structural similarities, motors can be
classified into two groups according to their function. Processive motors
\cite{howard97}, also called ``porters'' \cite{leibler93}, consist of a single
molecule which can move over long distances along its molecular track without
dissociating from it. The most common processive motor is kinesin interacting
with microtubules. Nonprocessive motors, also called ``rowers'', can only
generate macroscopic motion when operating in large groups. Muscular myosin,
interacting with actin, belongs to this class of motors. Here we focus on
nonprocessive motors.

For many decades exclusively data from physiological measurements on muscles
\cite{hill38} provided experimental information for modeling collective
molecular motors \cite{huxley57,huxley71}.  In recent years, a variety of {\it
in vitro\/} experimental techniques allowed the observation of single motor
proteins. These experiments include gliding assays \cite{toyoshima87,howard89},
optical tweezers \cite{svoboda93,finer94} and micromechanical force
measurements \cite{yanagida95}. They allowed for a new insight into the basic
principles underlying the operation of motors.  Not only new theoretical models
for single-molecule motors \cite{ajdari92,peskin95,duke96,derenyi96} were
inspired by these experiments, but also new models for cooperative motors
\cite{leibler93,prost95,juelicher97b}.  Except for the work by Csah{\'o}k et
al. \cite{csahok97}, which discusses the transport of elastically coupled
particles driven by colored noise, all these models deal with motors placed on
a rigid backbone, interacting with a rigid track. 

The assumption of stiff filaments seems to be appropriate under physiological
conditions in muscles, since the measured extensions of few nanometers
\cite{huxley94,wakabayashi94,higuchi95} are sufficiently small compared to the
myosin step-size, which is about $10\,{\rm nm}$ \cite{finer95}.  However, it
certainly can become invalid in gliding assays of the type discussed in Refs.\
\cite{toyoshima87,riveline98} if an actin filament glides over widely
separated linearly placed motors.  If the spacing between
motors is large enough, the elasticity of the backbone or track section between
two motors can become comparable to the elasticity of a single motor head.
Experiments with myosin molecules bound to an elastic background are
conceivable as well.

\begin{figure}
\centerline{
 \epsfxsize=0.9\columnwidth \epsfbox{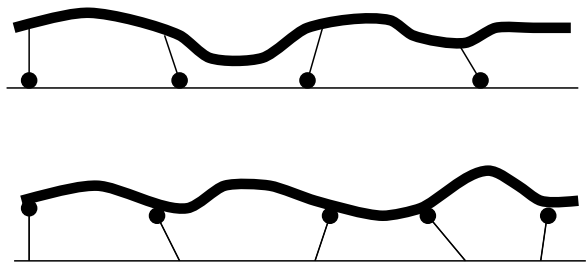}
}
\vspace*{0.5cm}
  \caption{Two possible sources of elastic coupling of collective molecular
  motors: elastic backbone (upper figure) and elastic track (lower
  figure). Except for very soft backbones/tracks, the description of both
  models is equivalent.}
  \label{fig1}
\end{figure}

It is the purpose of this paper to investigate qualitatively and quantitatively
the influence of filament elasticity on the operation of myosin-like
motors. Our major quantity of interest is the force-velocity-relation (filament
velocity as a function of the external load).  We want to identify the
universal effects of filament elasticity and at the same time keep the model as
close as possible to experiments.  The elasticity modeled by linear harmonic
springs may either originate from the flexible backbone or from the flexible
track (Fig.\ \ref{fig1}). As long as we are dealing with small relative
elongations, both sources of flexibility are equivalent and our model
should apply to both cases.  For specificity we use a formulation with an
elastic backbone and a stiff track.

\pagebreak

Another important component of the model is the modeling of individual motor
heads. Most models describe the heads by several states with different
conformations and transition rates between them. Such models are based on early
ideas by A.F. Huxley \cite{huxley57} and have later been refined in order to
explain more experimental details. More recently simplified models using a
two-state ratchet formalism have been developed in order to concentrate on
generic features of motion generation \cite{juelicher97b}. Another class of
models describing Brownian particles in ratchet potentials subject to colored
noise, however, seems to be only very distantly related to motor proteins
\cite{leibler94}.  We decided to use a two-state crossbridge model, similar to
the model introduced by Leibler and Huse \cite{leibler93}, but with just two
long-living states. It includes the transitions between attached and detached
state and an active power stroke, which have both been identified as basic
elements of the myosin motor \cite{huxley71,howard97} and also observed {\it in
  vitro} \cite{finer94}. Compared to other investigations
\cite{huxley57,leibler93,juelicher97b} our model also contains a low number of
free parameters, which makes it more suitable for a study of universal aspects.
Yet, adding the strain-dependence of the detachment rate, the model we use is
sufficient to describe the experimentally measured force-velocity-relation of
actin/myosin in muscles \cite{hill38}. A more detailed description of this
two-state model can be found in Ref.\ \cite{vilfan_preprint}.

Due to the generic features of such two-state models we expect that the effects
discussed here should as well apply to other models which contain the same
basic mechanisms of force generation, e.g.\ \cite{leibler93,juelicher97b}.
However, models of the type discussed in Ref.\ \cite{csahok97}, which describe
particles in an asymmetric periodic potential subject to a temporally
correlated noise, are based on a thoroughly different driving mechanism.
Therefore, one expects and actually finds a variety of disparate effects
including a strong influence of the coupling strength on the velocity even
without external load.

The outline of the paper is as follows. In Sec.\ \ref{sec2} we introduce the
model \cite{note_applet}, describe its phenomenological properties, present the
main results and discuss its implications to experiments.  The full calculation
for strain-independent detachment rates is shown in Sec.\ \ref{sec3} and for
strain-dependent rates in Sec.\ \ref{sec4}.

\section{Discussion}
\label{sec2}
\subsection{Description of the model}
\label{sec2.1}

\begin{figure}
\centerline{
 \epsfxsize=0.9\columnwidth \epsfbox{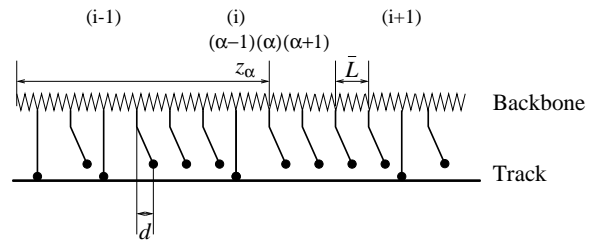}
}\vspace*{0.5cm}
  \caption{Definition of the model. Motors are fixed on the elastic backbone
  at uniform spacing $\bar L$ and attach to the stiff track. $z_\alpha$ denotes
  the position of $\alpha$-th motor on the unstrained backbone. Due to the
  conformational change, the head of each motor attaches at the distance $d$
  from its root.}
  \label{fig2}
\end{figure}

We consider a one-dimensional model describing many motors which produce force
between two filaments gliding past each other.  The force is generated by a
conformational change (``power stroke'') in the molecular motors fixed to the
{\it backbone}, which takes place after they attach to the molecular {\it
track\/} (Fig.\ \ref{fig2}). We assume that the motor proteins can be found in
two states: attached to and detached from the track. This corresponds to taking
into account only the two long-living states in the model used by Leibler and
Huse \cite{leibler93}. The transitions between these two states occur
stochastically.  We denote the mean life time of the attached state by $t_a$
and of the detached state by $t_d$.  Each attached motor is described as a
harmonic spring connecting its root at the backbone (position $y$) and its head
on the track (position $x$). The force this motor produces between the track
and the backbone is given as $k(x-y)$, with a spring constant $k$. Since the
motor is in a forward leaning position before attachment (see Fig.\
\ref{fig2}), it attaches to the track at the point $x^n$, which is the position
$y$ of the root of the motor before attachment, shifted by the displacement $d$
(``power stroke'')
\begin{equation}
x^n=y+d\;.
\end{equation}

We neglect thermal fluctuations and the discreteness of binding sites on the
track. This is motivated by our recent results for rigid backbones
\cite{vilfan_preprint}, where we find that discrete binding sites (with a
spacing of $5.5\,{\rm nm}$) as well as thermal fluctuations have only a minor
effect on the resulting force-velocity-relation.

\begin{figure}
\centerline{
 \epsfxsize=0.9\columnwidth
 \epsfbox{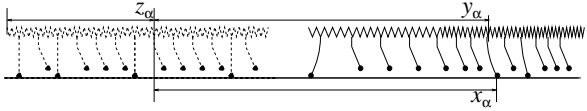}
}\vspace*{0.5cm}
  \caption{The moving backbone (solid line) and its initial position at time
    $t=0$ (dashed line). $y_\alpha$ denotes the root position of the
    $\alpha$-th motor, relative to its initial value ($z_\alpha$). $x_\alpha$
    denotes the head position of the $\alpha$-th motor, also relative to the
    initial position of the $\alpha$-th root, $z_\alpha$.}
  \label{fig3}
\end{figure}

While assuming a stiff track, we model the backbone as a linear spring with
compliance $\gamma^{-1}$ per unit length.  But note that this is merely a
convention. Our results apply equally well to the reverse situation, where the
molecular motors are fixed on a (rigid) cover slip and the elasticity is due to
the molecular track transported by them.  We consider $\bar N$ motors placed on
the backbone at uniform spacing $\bar L$, so that the total backbone length is
$\bar N \bar L$. Note that the assumption of uniform spacing is made solely for
simplicity. Any other distribution which is homogeneous on length scales $L$
would lead to the same exponential distribution of gap widths (see
Eq.~\ref{eq2.3} below). The position of the $\alpha$-th motor on the unstrained
backbone will be referred to as $z_\alpha$. In the following, the actual
positions of motor heads $x_\alpha(t)$ and of motor roots $y_\alpha(t)$ are
measured relative to $z_\alpha$ (Fig.\ \ref{fig3}).

Instead of using the quantities $\bar N$ and $\bar L$ it will prove helpful to
use the mean number of attached motors $N=\bar N {t_a}/{(t_a+t_d)}$ and the
mean spacing between two attached motors $L=\bar L {(t_a +t_d)}/{t_a}$.  Since
for now we are dealing with strain-independent reaction rates the distribution
of attached and detached states does not depend on the motion of their
positions.  The probability of finding a gap with $\alpha$ detached motors
between two attached ones is given by the geometric distribution
\begin{equation}
  \label{eq2.2}
p_\alpha=\left(\frac{t_d}{t_a +t_d}\right)^\alpha \frac{t_a}{t_a +t_d}\;.
\end{equation}

In order to keep the model as lucid as possible we assume a small duty ratio
\cite{howard97}, meaning that a motor molecule spends most of its time in the
detached state, $t_a \ll t_d$.  Since we are dealing with nonprocessive motors
(``rowers'' \cite{leibler93}), this assumption is certainly valid.  While
keeping the mean number $N$ of attached motors and their average spacing $L$
constant, we consider the limit $\bar L\to 0$. With this simplification the
model becomes continuous. Also, the assumption about equidistantly placed
motors on the backbone becomes superfluous in this limit. The distribution of
motors is insignificant as long as it is sufficiently homogeneous on the length
scale $L$.  The attachment rate per length $L$ (between the positions $z$ and
$z+L$ on the backbone) obeys $r_a= L/{\bar L t_d}=1/{t_a}$. The distribution of
gap widths (\ref{eq2.2}) takes the form of an exponential distribution
\begin{equation}
\label{eq2.3}
p(l)=\frac 1 L e^{-l/L}\;.
\end{equation}

\subsection{Results}

In this subsection we summarize our main results for the analysis of the model
described above; the details of the calculation are given later in Sec.\
\ref{sec3}.

As described by now the model contains seven independent parameters: $t_a$,
$N$, $L$, $F$, $k$, $\gamma$ and $d$. Upon measuring the force per motor,
$F/N$, in units of the force during one power stroke, $kd$, and measuring the
backbone elasticity per unit length, $\gamma/L$, in units of the motor head
elasticity, they may be reduced to two adimensional parameters: $\hat F=F/Nkd$
and $\hat\gamma=\gamma/kL$. Then the velocity in units of a single motor
velocity $\hat v =v t_a/d$ is given by a ``scaling'' function $\hat v=\eta(\hat
F, \hat\gamma)$.

\begin{figure}[tbh]
\centerline{
 \epsfxsize=0.9\columnwidth \epsfbox{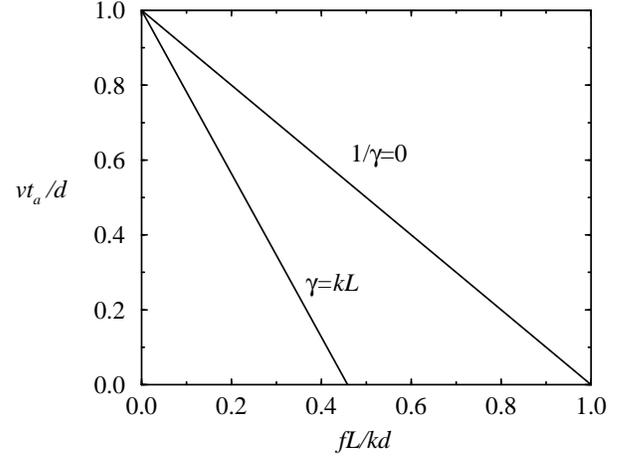}
}\vspace*{0.5cm}
  \caption{Force-velocity-relation for the stiff ($\gamma=\infty$) and elastic
  ($\gamma=kL$) backbone.}
  \label{fig7}
\end{figure}

As shown in Sec.\ \ref{sec3} we find that in case of a time-independent
external force it does not matter whether this force pulls on one end or
homogeneously on the whole backbone with a density $f=F/NL$. A quite remarkable
result of our analysis is that the force-velocity relation remains linear for
flexible backbones and that the zero load velocity $d/t_a$ does not depend on
the backbone elasticity. The force-velocity relations for a stiff and an
elastic backbone are shown in Fig. \ref{fig7}.

\begin{figure}[tbh]
\centerline{
 \epsfxsize=0.9\columnwidth \epsfbox{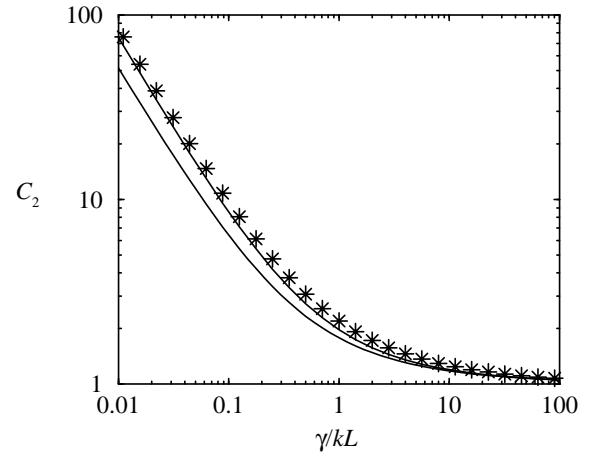}
}\vspace*{0.5cm}
  \caption{The slope of the force-velocity relation (coefficient $C_2$ in
  Eq.\ (\protect\ref{eq4.4})), as a function of the relative backbone stiffness
  $\gamma/kL$: MC-simulations (stars), analytic approximation without
  correlations (lower curve), with correlations to the distances to nearest
  neighbors (upper curve). In both analytic curves the numeric result for
  $k\langle 1/K_e\rangle$ from Fig.\ \protect\ref{fig6} was used.}
  \label{fig9}
\end{figure}

If the relative backbone stiffness $\hat\gamma$ lies below $1$, the slope of
the force-velocity curve and consequently the stall force differ significantly
from those for a stiff backbone (Fig.\ \ref{fig9}).  While the stall force
$f_{\rm stall}$ is proportional to the motor stiffness ($f_{\rm stall}= k d /
L$) for stiff backbones, it becomes a function of backbone stiffness for very
flexible backbones and is given as $2 \gamma d / \nu L^2$.  Here $\nu$ is a
numerical constant which has the value $\nu\approx 1.64$ as obtained from the
Monte-Carlo simulation. Using analytical tools based on a Master-Equation
approach with correlations between the position of a motor and the distances to
its neighbors, we obtain the value $\nu\approx 1.50$, which is in good
agreement.  For completeness Fig. \ref{fig9} also shows the result of a
Master-Equation without correlations, yielding $\nu=1$.

Finally, we extend the model described in Sec.\ \ref{sec2.1} by introducing a
strain-dependent detachment rate. This extension is inevitable for a
quantitative comparison with experiments on the actin-myosin system. We already
mentioned that (with or without backbone elasticity) strain-independent
transition rates lead to a linear force-velocity-relation. However, since the
very beginning of muscle research it has been known that the
force-velocity-relation rather has a hyperbolic form, also called the Hill
curve \cite{hill38}. It has also been known that the energy liberation in a
stretching muscle depends on velocity, which is also called the Fenn effect
\cite{fenn24}. The most natural way to reproduce these physiological
measurements is the introduction of a strain-dependent detachment rate, meaning
that the lifetime of the attached state $t_a=t_a(x_i-y_i)$ is larger for those
heads that have just gone through the power stroke and produce maximum force
than for those which have already done their work and now pull backwards.  As a
consequence the duty ratio becomes lower at higher velocities.  This idea has
already been used by A.F. Huxley \cite{huxley57}.  Although there is some
direct experimental evidence for strain-dependent detachment rates
\cite{finer94}, the functional form of this dependence has not been measured
yet.  For simplicity we model this dependence as an exponential
$t_a(\xi)=\exp(\alpha \xi)$, which suffices to fit the force-velocity-relation
\cite{hill38}.  However, we stress that this is only a first approximation and
that other forms are possible as well. Further experimental information is
highly desirable for a future more detailed modeling of molecular motors.

Some other functional forms of the detachment rate (e.g.\ $t_a(\xi)\propto
\xi^{-2}$) lead to anomalous force-velocity-relations already with rigid
backbones \cite{vilfan_preprint}. These can lead to oscillations, similar to
those proposed by J\"ulicher and Prost \cite{juelicher97}.  We expect that
flexible backbones can lead to additional phenomena like wave generation.

Strain-dependent detachment rates enhance the difficulty of an analytic
solution of our model enormously, since the distribution of attached and
detached motors depends on the distribution of head positions
($x_i$). Therefore, we will mainly use Monte-Carlo simulations and restrict
analytic arguments on limiting cases.  The simulations show that two major
analytic results of the strain-independent case carry over to the
strain-dependent case: (i) If the backbone flexibility exceeds its
characteristic value, the stall force decreases strongly. (ii) The backbone
flexibility has only little influence on the zero-load velocity.

\begin{figure}
\centerline{
 \epsfxsize=0.9\columnwidth \epsfbox{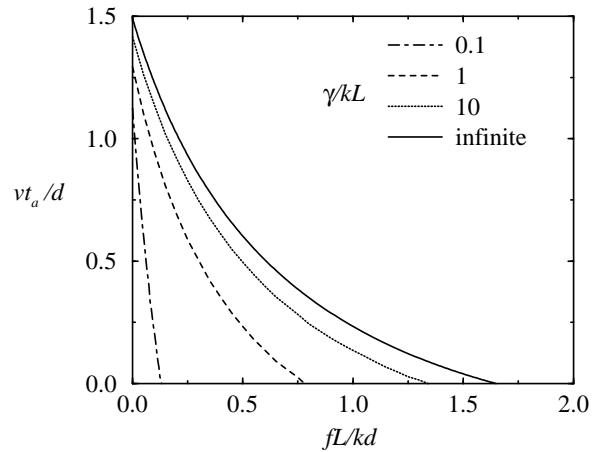}
}
  \caption{Monte-Carlo results for force-velocity-relations with
  strain-dependent detachment rates ($\alpha d=0.5$) and different backbone
  stiffnesses. Note the crossover from the Hill curve at stiff backbones to a
  nearly linear relation at soft backbones.}
  \label{fig11}
\end{figure}

For stiff backbones ($\gamma/kL\gg 1$) the force-velocity-relation as measured
by Hill \cite{hill38} is reproduced perfectly using the value $\alpha
d=0.58$. For decreasing backbone stiffness ($\gamma/kL\ll 1$) the stall force
decreases and the force-velocity-curve becomes increasingly linear. This can be
understood as follows: The forces which lead to positive velocities become
smaller and so does the strain on single motors. The small strain does not have
any significant influence on the detachment rate any more and the results
obtained for a strain-independent detachment rate become exact. The crossover
from the Hill curve to a linear relation is shown in Fig.\ \ref{fig11}. There
$\alpha d=0.5$ is used, but the curves would look qualitatively similar for
other positive values.

\subsection{Implications for experiments}

In order to apply our theory to experiments, we need the spring constants of
the myosin heads and of the actin filaments. The elasticity of the attached
myosin head was measured by Finer et al.\ \cite{finer95} with the result
$k=0.4\,{\rm pN}/{\rm nm}$ (measurements by Ishijima et al.\ \cite{ishijima96}
yield $k=0.28\,{\rm pN}/{\rm nm}$, which is in quite good agreement).

In our model we assumed springs with flexibility proportional to their length,
obeying $k_{\rm spring}=\gamma/L$. Actually actin is a semiflexible
polymer. Its elastic behavior was subject of many theoretical and experimental
studies in last years
\cite{mackintosh-kaes-janmey:95,kroy-frey:96,wilhelm-frey:96,frey-etal:condmat,kojima94}.
There are essentially two contributions to the elasticity of actin: the
longitudinal elastic modulus and the buckling of the polymer, induced through
thermal fluctuations.

At very high loads, the stiffness is limited by the elastic modulus of actin
and is proportional to $L^{-1}$. For a filament of length $1\,\mu{\rm m}$ the
stiffness is about $44\,{\rm pN}/{\rm nm}$ \cite{kojima94}. In this regime, the
characteristic distance between attached motors can be estimated as $L_{\rm
ch}=\gamma/k\approx 100\,\mu{\rm m}$. In this case backbone flexibility is of
no experimental relevance.

At low tension, the buckling modes limit the stiffness of actin. In the linear
response approximation, the stiffness of a polymer with length $L$ obeys the
law
\begin{equation}
\label{eq6.1}
k_{\rm polymer}=90 k_B T \ell_p^2L^{-4}
\end{equation}
where $\ell_p$ denotes the persistence length.  Although we assume springs
obeying an $L^{-1}$-law, we can still use the $L^{-4}$-law to give an estimate
for the characteristic distance for which $k_{\rm polymer}(L_{\rm ch})\approx
k$.

Recent measurements provide the value $\ell_p=7.4\mu{\rm m}$
\cite{riveline-etal:97}. With these values we finally estimate the
characteristic distance between attached motors $L_{\rm ch}=(90 k_B T \ell_p^2
/ k)^{1/4}\approx 500\,{\rm nm}$. If the mean displacement between attached
motors is larger than $L_{\rm ch}$, we expect that the effects of backbone
elasticity should be observable. A simulation with springs obeying the
$L^{-4}$-law (\ref{eq6.1}) is in preparation.

In muscles $500\,{\rm nm}$ is about the length of a half sarcomere
\cite{bagshaw93,alberts}. A rough estimate (300 myosin heads in one thick
filament, 3 actin filaments per one thick filament, $2.5-10\%$ of heads in the
attached state) leads to $L$ values between $50\,{\rm nm}$ and $200\,{\rm nm}$,
significantly below the characteristic length $L_{\rm ch}$. This implies that
the elasticity of actin filaments does not influence the operation of muscles.
This result is not surprising -- backbone elasticity always reduces the
efficiency of motors and it would be hard to understand why muscles spoil
their high efficiency in such a prodigal way.

\section{Analytical solution}
\label{sec3}

In this section the calculation leading to the force-velocity-relation for
strain-independent detachment rates is given. This is done in several steps:
first we show that the model behaves equivalently if a force acts on one end of
the filament or homogeneously along the whole length. As second we calculate
the effective compliance of a semi-infinite chain, which is an important input
quantity for later use. Then we show the linearity of the
force-velocity-relation, show that the zero-load velocity does not differ from
its value for stiff backbones and finally calculate its slope in two different
approximations.

A generic situation found in many experimental setups is that a force $F$,
usually produced by an optical tweezer, acts on the rear end of the
backbone. Another possibility is to produce the force by a viscous liquid,
flowing along the backbone. Such a force acts more or less homogeneously on the
whole backbone.  In both cases the main quantity of interest is the resulting
mean backbone velocity depending on the load $F$.  In the following we show
that both situations are equivalent within the scope of a theoretical
description.

\subsection{Point force}
From now on we use the index $i$, which runs over attached motors only, instead
of the index $\alpha$, running over all motors.  $x_i$ and $y_i$ denote the
head and root positions of $i$-th attached motor relative to its initial
position on the unstrained backbone ($z_i$) at time $t=0$.  The stiffness of
the backbone fragment between the motors $i$ and $i+1$ equals
$\gamma/(z_{i+1}-z_i)$. At the point where the $i$-th motor is fixed to the
backbone, the sum of the all three forces (from the motor, from the left part
of the backbone and from the right part of the backbone) must be zero:
\begin{mathletters}
\label{eq2.4}
\begin{eqnarray}
k(x_i-y_i)-\gamma\frac{y_i-y_{i-1}}{z_i-z_{i-1}}+\gamma\frac{y_{i+1}-y_i}{z_{i+1}-z_i}&=&0\;,\\
k(x_1-y_1)+\gamma\frac{y_{2}-y_1}{z_{2}-z_1}&=&F\;.
\end{eqnarray}
\end{mathletters}
The second equation describes the first attached motor and differs from the
others since the backbone force acting from the left is replaced by the
external force $F$. With given $x_i$ and $z_i$ this set of equations allows us
to determine the values of $y_i$.

The {\it detachment\/} rate  equals  $t_a^{-1}$ for each motor.  The
detachment of one motor is described by canceling its position in the set of
$x$- and $z$-values. Afterwards, all the $y$-values are determined anew from
Eq.\ (\ref{eq2.4}).

The process of {\em attachment\/} occurs at the rate $N/t_a$ and consists of
choosing a random position $z^n$ between $0$ and $NL$, calculating the
corresponding $y(z^n)$ (the root position of the new motor before
attaching) and $x^n=y(z^n)+d$, and finally adding a new motor with its head at
$x^n$ and its root at $z^n$.  Again, all the $y$-values have to be recalculated
as stated by Eq.\ (\ref{eq2.4}).  Expressing $y(z^n)$ through the positions of
the neighbors (the index ``$-$'' describes the first attached motor on the left
and ``$+$'' on the right hand side) yields
\begin{equation}
\label{eq2.5}
x^n=y^n+d=\frac{y_-(z_+-z^n)+y_+(z^n-z_-)}{z_+-z_-}+d\;.
\end{equation}

\subsection{Equivalence to the model with a homogeneous force}

In the model  described by now, the external force acts on one end of
the backbone.  This leads to some difficulties, e.g.\ one can only consider a
semi-infinite chain with one boundary condition.  Furthermore, the resulting
solutions are not translationally invariant since the strain decreases along
the backbone.

Replacing the point force by a homogeneous one acting on the whole
backbone with a density $f={dF}/{d z}$ would allow us to perform the
calculation on an infinite chain of completely equivalent motors.  The ability
to use periodic boundary conditions in the Monte-Carlo-simulation would be an
additional advantage.

Fortunately, both models, i.e.\ with a point force and a homogeneous force are
actually equivalent. It is instructive to show this equivalence in the
continuum formulation of the model. Instead of the discrete set of variables
$y^P_i$ we use a function $y^P(z)$. The continuum representation of Eqn.\
(\ref{eq2.4})) is given by the following set of equations
\begin{mathletters}
\label{eq2.6}
\begin{equation}
\gamma \frac{d^2y^P}{dz^2} = - \sum_i \delta(z-z_i)
k(x^P_i-y^P(z_i))\;,
\end{equation}
\begin{equation}
\gamma\left. \frac {dy^P}{dz}\right|_{z=0}=F\;.
\end{equation}
\end{mathletters}
The first equation expresses the constant tension between the attached motors
with jumps at the positions where the motor roots are placed.  The second
equation describes the strain at the boundary of the backbone.

On the other hand, in the homogeneous force model the strain grows linearly
with $z$
\begin{equation}
\label{eq2.7}
\gamma \frac{d^2y^H}{dz^2} =f -\sum_i \delta(z-z_i) k(x^H_i-y^H(z_i))\;.
\end{equation}
\begin{figure}
\centerline{
 \epsfxsize=0.9\columnwidth \epsfbox{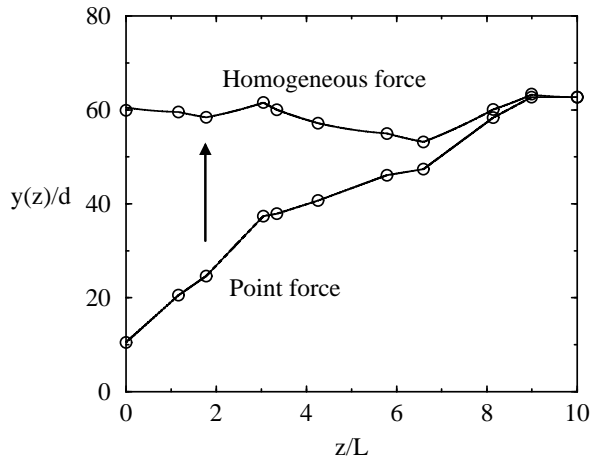}}
  \caption{Transformation from the model with a point force to the model with
  a continuous force, given through Eq.\ (\protect\ref{eq2.8}), on a typical
  configuration during the motion. The circles
  represent attached motors, the detached ones lie on the line. $z$ denotes
  the position a motor would have on an unstrained backbone, $y$ its root
  position relative to $z$ (Fig.\ \protect\ref{fig3}).}
  \label{fig4}
\end{figure}
Since we are dealing with a force constant in time and with quasistationary
solutions, we can show that both models are equivalent up to a transformation
which shifts the heads and roots of motors, depending on their $z$ position.
As may be easily verified by comparing Eqns.\ (\ref{eq2.6}) and (\ref{eq2.7}),
the following transformation
\begin{equation}
\label{eq2.8}
\left\{ x_i^{H} \atop y_i^{H} \right\}=\left\{x_i^P \atop y_i^P\right\}+\frac{f}{2 \gamma} (NL-z_i)^2
\end{equation}
with $F=N L f$ preserves the properties of the model. The transformation is
shown schematically in Fig.\ \ref{fig4}.  After having shown the
equivalence of both models, we can return to the original formulation. The
transformed equations (\ref{eq2.4}) and (\ref{eq2.5}) become (in the following we
omit the index ${}^H$)
\begin{equation}
\label{eq2.9}
k(x_i-y_i)-\gamma\frac{y_i-y_{i-1}}{z_i-z_{i-1}}+
\gamma\frac{y_{i+1}-y_i}{z_{i+1}-z_i}-f\frac{z_{i+1}-z_{i-1}}{2}=0
\end{equation}
and
\begin{eqnarray}
\label{eq2.10}
x^n&=&\frac{y_-(z_+-z^n)+y_+(z^n-z_-)}{z_+-z_-}\nonumber\\&&+d-
\frac{f}{2\gamma}(z_+-z^n)(z^n-z_-)\;.
\end{eqnarray}
The additional term represents the displacement of a uniformly loaded spring,
tightly bound at its ends at $z_-$ and $z_+$.  Of course, these equations also
follow directly from (\ref{eq2.7}).

\subsection{Effective compliance of a semi-infinite chain}
\label{sec3.3}

\begin{figure}
\centerline{
 \epsfxsize=0.9\columnwidth \epsfbox{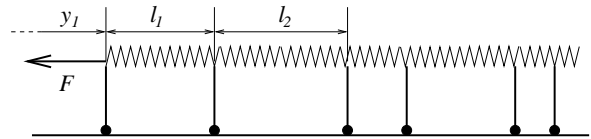}
}\vspace*{0.5cm}
  \caption{Effective spring constant of a semi-infinite chain with randomly
  distributed displacements between bridges.}
  \label{fig5}
\end{figure}
A quantity frequently needed during the analytical solution of the model
described in Sect.\ \ref{sec2} is the elasticity of a semi-infinite
stochastic chain as shown in Figure \ref{fig5}.  It is defined as
\begin{equation}
\frac 1 {K_e}=-\left. \frac{d y_1}{dF} \right|_{x_i = {\rm const}}
\end{equation}
where $y_1$ is a part of the solution of Eq.\ (\ref{eq2.4}).  As defined in
Sect.\ \ref{sec2.1}, the spring constant of the motors is given by $k$. The
values of $l_i$ are distributed randomly with average $L$ and the exponential
distribution (\ref{eq2.3}).

\begin{figure}[bth]
\centerline{
 \epsfxsize=0.9\columnwidth \epsfbox{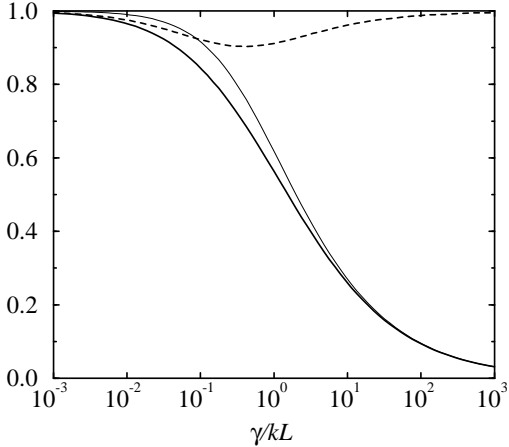}
}\vspace*{0.5cm}
  \caption{The effective compliance of a semi-infinite chain 
  $k\langle 1/K_e\rangle$.  Thick line: Monte-Carlo result.  Thin line:
  approximation given by Eq.\ (\protect\ref{eq3.5}). Dashed line: the ratio
  between the Monte-Carlo result and the approximation.}
  \label{fig6}
\end{figure}

The compliance of a chain with a given configuration (given $l_i$-values) can
be calculated recursively. The chain is built up of a spring with stiffness
$k$, connected in parallel to two other springs, which themselves are connected
in series. The first one describes the piece of backbone with elasticity
$\gamma/l_1$. The second spring is again a replacement for another
semi-infinite chain starting with the second motor. We denote its stiffness as
$K_e^\prime$.
\begin{equation}
K_e=k+\frac 1 {\frac {l_1}{\gamma} +\frac {1}{K_e^\prime}}
\end{equation}
Repeating the same procedure for $K_e^\prime$ etc.\ and finally averaging over
all configurations $(l_1,l_2,\ldots)$ with their statistical weights yields
\begin{equation}
\label{eq3.3}
\left< \frac{1}{K_e} \right>=\int\limits_0^\infty dl_1 p(l_1)
\int\limits_0^\infty dl_2 p(l_2) \ldots \frac1{k+\frac1{\frac{l_1}{\gamma}+
 \frac1{k+\frac1{\frac{l_2}{\gamma}+ 
 \frac1{k+\frac1{\cdots}}}}}}\;.
\end{equation}
The convenient way to solve this high-dimensional integral, however, is by
using the Monte-Carlo method. Its result is shown in Fig.\ \ref{fig6}.

One possibility to give an analytic approximation for $\left< K^{-1}_e \right>$
is the use of a mean-field like theory by assuming that all displacements
$l_i$ are exactly equal to their mean value $L$.  Then the above expression
becomes self-similar and we obtain the recursion relation
\begin{equation}
\left< \frac{1}{K_e} \right>=
 \frac1{k+\frac1{\frac{L}{\gamma}+\left< \frac{1}{K_e} \right>}}
\end{equation}
with the solution
\begin{equation}
\label{eq3.5}
\left< \frac{1}{K_e} \right>=
\frac{L}{2\gamma}\left(\sqrt{1+4\frac{\gamma}{kL}}-1\right)\;.
\end{equation}
In Fig.\ \ref{fig6} the approximation, Eq.\ (\ref{eq3.5}) is plotted
against the (exact) Monte-Carlo solution of (\ref{eq3.3}).  The deviations lie
below $10\%$ over the entire parameter range.  The approximation becomes exact
for both very soft and very rigid backbones.  In the case of large $\gamma$
this is due to the fact that the long-range coupling makes the detailed
distribution of $l_i$ irrelevant. The case of low $\gamma$ is trivial since the
motors get decoupled and the stiffness of the chain is determined solely by the
first motor ($\left< K^{-1}_e \right>=\frac 1 k$).

\subsection{Linearity of the force velocity relation}

For rigidly coupled motors it was already shown in \cite{leibler93} that the
velocity is linear in the applied force $f$ as well as in the step size $d$.
In the following we present a simple argument why this remains valid for
elastically coupled motors.  We denote the number of currently attached motors
by $n$ and their positions on the backbone by a vector ${\bf x}$ containing the
components $x_1\ldots x_n$.  From the structure of Eqns.\ (\ref{eq2.9}) and
(\ref{eq2.10}) it is evident that the process of attachment can be described by
the following equation
\begin{equation}
{\bf x}^{(i+1)}=A{\bf x}^{(i)}+{\bf u}_1 \frac {fL} k + {\bf u}_2 d
\end{equation}
where the $(n+1,n)$-matrix $A$ and the $(n+1)$-vectors ${\bf u}_1$ and ${\bf
u}_2$ depend in a complex way on $\{z_i\}$ and $\gamma/Lk$.  The detachment
of one head is described by another $(n-1,n)$-matrix A as
\begin{equation}
{\bf x}^{(i+1)}=A\bf{x}^{(i)}\;.
\end{equation}
After a series of consecutive attachments and detachments this gives 
\begin{equation}
{\bf x}^{(i)}=\tilde A{\bf x}^{(0)}+{\bf \tilde u}_1 \frac {fL} k + {\bf \tilde
u}_2 d\;.
\end{equation}
Finally we set ${\bf x}^{(0)}=0$ and calculate the mean motor position $\left<
x \right>={\rm tr}\,{\bf x}/n$.  Since the time needed for $i$ steps is
proportional to $t_a$, we obtain the relation
\begin{equation}
\label{eq4.4}
v=\frac 1{t_a} \left(C_1 d-C_2 \frac {f L} k \right)\;,
\end{equation}
which is linear in $d$ and in $f$.  The constants $C_1$ and $C_2$ get
independent of the mean number of motors $N$ for $N\to \infty$.

The force-velocity-relations for the elastic and for the stiff backbone are
compared in Fig.\ \ref{fig7}.  Due to the linearity of the force-velocity
relation the problem can be separated in two parts: determining the backbone
velocity without external forces ($f=0$) and the velocity with external force
but without power strokes ($d=0$).  The remaining work consists of determining
the constants $C_1$ and $C_2$.

\subsection{Master-equation}

An adequate description of the model is given by $P(\ldots,l_2,l_1 ; x,t ;
r_1,r_2,\ldots)$, the probability density to find a head at position $x$ and
with distances $l_1\equiv z_i-z_{i-1}$ and $r_1\equiv z_{i+1}-z_i$ to its
nearest attached neighbors, the distances $l_2\equiv z_{i-1}-z_{i-2}$ and
$r_2\equiv z_{i+2}-z_{i+1}$ between the nearest and the next nearest attached
neighbors etc.. Of course, this distribution varies with time.  Because the
problem is linear in $x$ and $y$, there is no need for determining the
correlations between the positions of different motor heads on the track.
$P$ can be expressed by
\begin{eqnarray}
\label{eq4.5}
&&P(\ldots,l_2,l_1 ; x,t ;
r_1,r_2,\ldots)=\nonumber\\&&P(x,t)_{\ldots,l_2,l_1 ; r_1,r_2,\ldots} p(\ldots,l_2,l_1 ;
r_1,r_2,\ldots)\;.
\end{eqnarray}
The second factor describes the probability for a motor to have the distances
$l_1$, $r_1$, $l_2$, $r_2,\ldots$ to its neighbors.  In a steady solution it is
given by Eq.\ (\ref{eq2.3})
\begin{equation}
\label{eq4.6}
 p(\ldots,l_2,l_1 ; r_1,r_2,\ldots)= \frac{e^{-l_1/L}}{L} \frac{e^{-r_1/L}}{L}
\frac{e^{-l_2/L}}{L} \ldots 
\end{equation}
The first factor gives the distribution of the head positions $x$ for the given
set of distances.  Since the transition rates are constant
(strain-independent), the first factor does not influence the second one.

We describe the temporal development of $P$ in terms of a Master equation.
The detachment of motors is described as drain, attachment as source.
Additional terms result from the fact that the attachment/detachment of a motor
also changes the $l$ and $r$-values in its neighborhood.  
\begin{figure}
\centerline{
 \epsfxsize=0.9\columnwidth \epsfbox{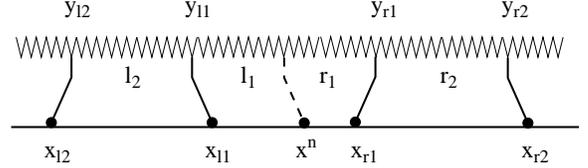}
}\vspace*{0.5cm}
  \caption{Attachment of a new head.}
  \label{fig8}
\end{figure}
\endmcols

The detachment rate is equal to the probability density divided through the
mean life time of the attached state
\begin{equation}
\label{eq4.7}
r_d(\ldots,l_2,l_1 ; x,t ; r_1,r_2,\ldots)=\frac 1 {t_a} P(\ldots,l_2,l_1 ; x,t ;
r_1,r_2,\ldots)\;.
\end{equation}
Once the distribution of $z$-values ( $p(\ldots,l_2,l_1 ; r_1,r_2,\ldots)$) is
in equilibrium, the attachment and detachment rate integrated over $x$ have to
be equal. They can only differ in the $x$-dependence. Thus we write the
attachment rate the same way as the detachment rate except for a different
factor containing the distribution of $x$-positions of the newly attached
heads.
\begin{equation}
\label{eq4.8}
r_a(\ldots,l_2,l_1 ; x,t ; r_1,r_2,\ldots)=\frac 1 {t_a}  p(\ldots,l_2,l_1 ; r_1,r_2,\ldots) P^n(x,t)_{\ldots,l_2,l_1 ; r_1,r_2,\ldots}
\end{equation}
The $x$-distribution of attaching motors depends on the $x$-distributions of
all the neighbors. It is determined as the integral over all (properly
weighted) configurations which lead to a motor attachment at position
$x^n$:
\begin{equation}
\label{eq4.9}
P^n(x,t)_{\ldots,l_2,l_1 ; r_1,r_2,\ldots}=\ldots \int dx_{l1}
P(x_{l1},t)_{\ldots,l_3,l_2;l_1,r_1\ldots} \int dx_{r1}
P(x_{r1},t)_{\ldots,l_1,r_1;r_2,r_3\ldots} \ldots
\delta\left(x^n(\ldots,x_{l1},x_{r1},\ldots)-x\right)
\end{equation}
\beginmcols
As follows form Eq.\ (\ref{eq2.10}), $x^n$ can be expressed through $y_{l1}$ and
$y_{r1}$
\begin{equation}
\label{eq4.10}
x^n=\frac{y_{l1}r_1+y_{r1}l_1}{l_1+r_1}+d-\frac{f}{2\gamma} lr\;,
\end{equation}
which again are functions of all $x_i$ and can be determined through Eq.\
(\ref{eq2.9}).  The full Master equation is given in Appendix \ref{appendix1}.

\subsection{Zero load backbone velocity}

In the special case of zero external load ($f=0$) one can see that as long as
the expectation value $\langle x \rangle_{\ldots,l_2,l_1 ; r_1,r_2,\ldots}$ is
independent of the distances $l_i$ and $r_i$, i.e.\
\begin{equation}
  \label{eq4.11}
  \int dx\, x\, P(x,t)_{\ldots,l_2,l_1 ; r_1,r_2,\ldots}=\left< x
\right>\;,
\end{equation}
the same holds for $y_i$ (determined from (\ref{eq2.9})) and for $x^n$, which
follows from Eq.\ (\ref{eq4.10})
\begin{equation}
\label{eq4.12}
\left< x^n \right> =\left< y \right>+d=\left< x \right> +d\;.
\end{equation}
In other words, if the average head position of the existing attached motors is
not correlated to the distances between them, the position where the head of a
new motor attaches is uncorrelated too. Thus we have shown self-consistently
that $\langle x \rangle_{\ldots,l_2,l_1 ; r_1,r_2,\ldots}$ does not depend on
$l_i$ and $r_i$.  The equation of motion (\ref{eq_a2}) for the expectation
value of $x$ simplifies to
\begin{equation}
\label{eq4.13}
v\equiv \frac d {dt} \left< x \right>=\frac 1 {t_a} \left( \left< x^n \right> -
\left< x \right>\right) = \frac d {t_a}\;.
\end{equation}
This means 
\begin{equation}
C_1=1
\end{equation}
for the first coefficient in Eq.\ (\ref{eq4.4}).  The interesting point in this
result is that it is independent of the backbone elasticity.  As long as there
is no external force, the velocity remains the same as in the case of a rigid
backbone, this is $v(f=0)=d/{t_a}$. 

\subsection{Slope of the force-velocity curve (correlations neglected)}

In the previous section we have shown that in the case of zero external force
the average $x$-position of a motor stays uncorrelated to the distances to its
attached neighbors, which made the calculation of the force-velocity relation
pretty easy.  In the case of nonvanishing external forces (behavior described
by the coefficient $C_2$ in Eq.\ \ref{eq4.4}), the correlation doesn't vanish any
more.  However, as a first approximation we may still try to neglect it.  Later
we will take correlations with distances to the nearest neighbors into account
and show that they improve the result significantly.

The solution is analog to the previous section, except that we set $d=0$ and
$f\ne 0$ in Eq.\ (\ref{eq4.10}). Another difference is that 
the average $y$-value differs from the average $x$-value by the mean force per
attached motor ($fL$), divided through the motor stiffness $k$
\begin{equation}
\label{eq4.15}
\langle y \rangle= \langle x \rangle-\frac{fL}{k}\;.
\end{equation}
Instead of this rather intuitive argument this equation can also be derived
directly from Eq.\ (\ref{eq2.9}) by summation over $i$.

This equation describes the $y$-position, averaged over all motors. However,
the quantities needed in Eq.\ (\ref{eq4.10}) are the expectation values of
$y_{r1}$ and $y_{l1}$ (motors at the edge of a gap whose distance to one
neighbor is $l_1+r_1$ and to the other neighbor randomly distributed).  Even
with uncorrelated $x$-values the expectation values $\langle y_{l1} \rangle$
and $\langle y_{r1}\rangle$ are not the same as $\langle y \rangle$.

For a gap width equal to the mean distance between attached motors both
quantities will not differ.  Otherwise
the average value of the
$y$-positions beneath a gap of width $l_1+r_1$ differ
from $\left< y \right>$ by the excess force acting on the gap between them
($f(l_1+r_1)$) compared to the force on the average gap ($fL$), multiplied by
the effective compliance of a semi-infinite chain (Sect.\ \ref{sec3.3})
\begin{equation}
\label{eq4.16}
\langle y_{l1} \rangle = \langle y_{r1} \rangle = \langle y \rangle-f 
\left< \frac{1}{K_e} \right>  \frac{l_1+r_1-L}{2}
\end{equation}
The denominator $2$ describes the fact that the force is distributed equally to
both edges.

The mean $x$-position of the newly attached motor follows by combining Eqns.\
(\ref{eq4.10}), (\ref{eq4.15}), and (\ref{eq4.16}):
\begin{equation}
\label{eq4.17}
\langle x^n \rangle_{l_1;r_1} =\langle x \rangle-f \left( \left<
\frac{1}{K_e} \right> \frac{l_1+r_1-L}{2}+\frac{L}{k}+\frac{l_1
r_1}{2\gamma}\right)
\end{equation}

Since we want to neglect correlations between $\left< x \right>$ and
$(l_i,r_i)$, we average over $l_1$ and $r_1$ ($\left< l_1 \right>=\left< r_1
\right>=L$)
\begin{equation}
\langle x^n \rangle=\langle x \rangle-f \left( \left< \frac{1}{K_e} \right>
\frac{L}{2}+\frac{L}{k}+\frac{L^2}{2\gamma}\right)\;.
\end{equation}
In analogy to the previous section the final result for the dimensionless
coefficient $C_2$ is 
\begin{equation}
C_2= \left( \frac k 2 \left< \frac{1}{K_e} \right>
+1+\frac{k L}{2\gamma}\right)\;.
\end{equation}
The result is shown in Figure \ref{fig9}.  In the limit of very soft backbones
($\gamma \ll kL$) the stall force becomes $f_{\rm stall}=2d\gamma/L^2$, which
is independent of the motor stiffness $k$. Our Monte-Carlo simulations show
this behavior, however with a different prefactor, which results from the
correlations which were neglected in this approximation,
\begin{equation}
\label{eq4.20}
f_{\rm stall}=\frac{2d\gamma}{\nu L^2}
\end{equation}
with $\nu\approx 1.64$.

\subsection{First-order correlations}
\begin{figure}
\centerline{
 \epsfxsize=0.9\columnwidth \epsfbox{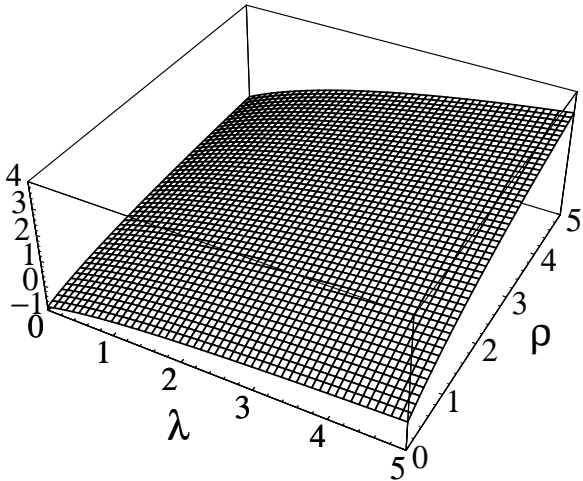}
}\vspace*{0.5cm}
  \caption{The function $\mu(\lambda;\rho)$ at $\gamma/kL=1$.}
  \label{fig10}
\end{figure}

Contrary to the previous subsection where we neglected the correlation between
the position $x$ of a motor and the distances to its neighbors by using the
ansatz (\ref{eq4.11}), we now extend the calculation by taking correlation with
distances to nearest neighbors into account.  We replace the approximation
(\ref{eq4.11}) by introducing a function describing these correlations
\begin{equation}
\int dx\, x\, P(x,t)_{\ldots,l_2,l_1 ; r_1,r_2,\ldots}=\left< x(t)
\right>-\frac{fL}{k}\mu\left(\frac{l_1}{L};\frac{r_1}{L}\right)\;.
\end{equation}
The function $\mu$ is scale invariant. It has to fulfill the condition
\begin{equation}
\int dl_1 \int dr_1 p(l_1;r_1) \mu\left(\frac{l_1}{L};\frac{r_1}{L}\right)=0\;.
\end{equation}
This also means that the head positions of the motors limiting a gap of width
$l_1+r_1$ already differ from $\langle x \rangle$
\begin{equation}
  \left< x \right>_{l_1+r_1;}=\left< x \right> -\frac{fL}{k} \int_0^\infty d
  r_2\, p(r_2) \mu\left(\frac{l_1+r_1}{L};\frac{ r_2}{L}\right)
\end{equation}
The equation (\ref{eq4.16}) has to be extended by a term describing this
influence. Since the roots of the motors are connected to their heads via
elastic constants $k$ and to the rest of the track via effective constants
$K_e -k $, the correction in $y$ corresponds to the $x$-correction attenuated
by the factor $k / K_e$. Finally, the refined Eq. (\ref{eq4.17}) reads
\endmcols
\begin{equation}
\langle x^n \rangle_{l_1;r_1} =\langle x \rangle-f \left( \left< \frac{1}{K_e}
\right> \frac{l_1+r_1-L}{2}+\frac{L}{k}+\frac{l_1 r_1}{2\gamma}\right) + k
\left< \frac{1}{K_e} \right> \left(\left< x \right>_{l_1+r_1;}-\left< x \right>
\right)\;.
\end{equation}
The correlation to farther neighbors is still neglected. This is expressed in
the simplified equation of motion which is obtained by integrating both sides
of (\ref{eq_a2}) over all $\lambda_{i\ge 2}$ and $\rho_{i\ge 2}$
\begin{eqnarray}
\label{eq4.25}
\lefteqn{\frac d{dt}\left< x(t) \right> p(\lambda_1;\rho_1) = \frac 1{t_a} \int
dl_1 \int dr_1 \ldots p(\ldots,l_1;r_1,\ldots)}\\ \bigl[&-&(\left< x(t)
\right>-(fL/k)\mu\left({l_1}/L;{r_1}/L\right))
\left(\delta(\lambda_1-l_1)\delta(\rho_1-r_1)\right)\nonumber\\
&+&(\left<x(t)\right>-(fL/k)\mu\left({l_2}/L;{l_1}/L\right))
(-\delta(\lambda_1-l_2) \delta(\rho_1-l_1)+\delta(\lambda_1-l_2)
\delta(\rho_1-(l_1+r_1)))\nonumber\\
&+&(\left<x(t)\right>-(fL/k)\mu\left({r_1}/L;{r_2}/L\right))
(-\delta(\lambda_1-r_1)\delta(\rho_1-r_2)+\delta(\lambda_1-(l_1+r_1))
\delta(\rho_1-r_2))\nonumber\\ &+&\left<x^n(t)\right>_{l_1;r_1}
\left(\delta(\lambda_1-l_1)\delta(\rho_1-r_1)\right)\nonumber\\
&+&(\left<x(t)\right>-(fL/k)\mu\left({l_2}/L;({l_1+r_1})/L\right)
(-\delta(\lambda_1-l_2) \delta(\rho_1-(l_1+r_1))+\delta(\lambda_1-l_2)
\delta(\rho_1-l_1))\nonumber\\
&+&(\left<x(t)\right>-(fL/k)\mu\left(({l_1+r_1})/L;{r_2}/L\right))
(-\delta(\lambda_1-(l_1+r_1)) \delta(\rho_1-r_2)+\delta(\lambda_1-r_1)
\delta(\rho_1-r_2))\bigr]\;,\nonumber
\end{eqnarray}
which leads to an integral equation for $\mu$ and $v$. Its scale invariant
form, using the coefficient $C_2=vt_a k / f$, is
\begin{eqnarray}
C_2&=&-3 \mu(\lambda;\rho)+ \int\limits_0^{\rho}
\mu(\lambda;\alpha)d\alpha+ \int\limits_0^{\lambda} 
\mu(\alpha;\rho)d\alpha+\left(k \left< \frac{1}{K_e} \right> \left(
\frac{\lambda+\rho-1}{2} \right) +1+\frac{\lambda
\rho}{2} \frac{kL}{\gamma}\right)\nonumber \\&&+ k \left< \frac{1}{K_e} \right>
\int\limits_0^\infty e^{-\alpha} \mu(\lambda+\rho;\alpha)d\alpha-(\lambda+\rho) \mu(\lambda;\rho) +\int\limits_0^\infty 
e^{-\alpha} \mu(\lambda;\rho+\alpha)d\alpha +\int\limits_0^\infty 
e^{-\alpha} \mu(\lambda+\alpha;\rho)d\alpha\;.
\end{eqnarray}
A self-consistent solution that holds for all $\lambda$ and $\rho$ can be
calculated numerically.  The resulting coefficient $C_2$ is shown in Fig.\
\ref{fig9} and a typical shape of the $\mu$-Function in Fig.\ \ref{fig10}.  The
correction $\nu$, defined in (\ref{eq4.20}), gets the value $\nu\approx 1.50$.
Taking first-order correlations into account improved the agreement between
theory and simulation significantly. The remaining deviation, due to neglected
correlations with further neighbors, is about $9\%$ at low $\gamma$ and lies
below $14\%$ over the entire parameter range.
\beginmcols

\section{Strain-dependent detachment rates}
\label{sec4}

\subsection{Stiff backbone}

A backbone with infinite stiffness ($\gamma\to \infty$) means that the motor
roots always keep their relative positions to each other, $y_i \equiv y$ for
all motors $i$. In this case the positions of motor roots $z_i$ naturally play
no role. The Master equation (\ref{eq_master}) for the $z$-independent
probability density $P(x,t)$ reduces to
\begin{equation}
\label{eq5.1}
\frac{d}{dt}P(x,t)=-r_d(x)+r_a(x)
\end{equation}
with $r_d(x)=P(x)/t_a(x-y)$, $r_a(x)= \delta(x-y-d)L/\bar L t_d$, as follows
from Eqns.\ (\ref{eq4.7}-\ref{eq4.10}).  The backbone position $y$ is
determined from Eq.\ (\ref{eq2.9}), summed over $i$:
\begin{equation}
\label{eq5.2}
k \int (x-y) P(x) dx= f L
\end{equation}
Note that the norm of the distribution $P$ gives the mean number of attached
motors per length $L$, which, contrary to previous sections, is not necessarily
equal $1$.  Quasistationary solutions of (\ref{eq5.1}) are found with
the ansatz
\begin{equation}
P(x,t)=\Phi (x-y)\;,\qquad y=vt\;,
\end{equation}
leading to
\begin{equation}
-v \partial_\xi \Phi(\xi)=-\frac{1}{t_a(\xi)} \Phi(\xi) + \frac{L}{\bar L
t_d} \delta(\xi-d)\;.
\end{equation}
This equation is analytically solvable. For $v>0$ the non-divergent solution
is
\begin{equation}
\Phi(\xi)=\frac{L}{v \bar L t_d} \exp\left(-\int_\xi^d \frac{d\xi'}{v t_a
(\xi')}\right)\theta(d-\xi)\;.
\end{equation}
Finally, the force-velocity relation can be obtained through (\ref{eq5.2})
$f=k\int \xi \Phi(\xi)d\xi/L$. The stall force is given as
\begin{equation}
f(v=0)=k d \frac{t_a(d)}{\bar L t_d}
\end{equation}
For simplicity reasons, we model the strain-dependence of the detachment rate
as an exponential function 
\begin{equation}
t_a(\xi)=t_a \exp(\alpha \xi)\;.
\end{equation}
The form of the
force-velocity-relation, which was first measured by Hill \cite{hill38}, is
fitted perfectly with $\alpha d=0.58$. Therefore, we use $\alpha d=0.5$ in the
following. 

\subsection{Elastic backbone}

In Section \ref{sec3} we showed that for very flexible backbones
($\gamma/kL\ll 1$) and strain-independent reaction rates ($\alpha=0$) the
zero-load velocity remains $\frac d {t_a}$, while the stall force is limited
through the backbone stiffness as $f(v=0)=2 \gamma d / \nu L^2$. The mean
strain on a motor is given as $\left< \xi \right> = {fL}/{k}$, or $2\gamma d /
\nu kL$ at maximum load. For very flexible backbones this means $\alpha
\left<\xi\right> \ll 2\alpha d \approx 1$. The strain-dependence becomes
negligible and the results from Sec.\ \ref{sec3} are exact. From this
simple argument we expect that the force-velocity relation becomes linear.  The
crossover from the hyperbolic to the linear shape takes place at
$\gamma/kL\approx 1$. While the stall force (force at zero velocity) $f_{\rm
stall}=e^{\alpha d} k d / L$ is limited through the motor stiffness $k$ for
stiff backbones, it depends solely on the backbone stiffness for soft
backbones, $f_{\rm stall}= 2 \gamma d / \nu L^2$. This behavior is in agreement
with our Monte-Carlo simulations, the results are shown in Fig.\
\ref{fig11}.

\section*{Acknowledgments}

We are grateful to Klaus Kroy, Rudolf Merkel and Erich Sackmann for helpful
discussions. This work has been supported by the Deutsche
Forschungsgemeinschaft under contract no.\ SFB 413.  A.V. and E.F. would like
to acknowledge support from the Cusanuswerk and by a Heisenberg fellowship from
the Deutsche Forschungsgemeinschaft, respectively.

\endmcols
\appendix

\newpage
\section{Master-equation}
\label{appendix1}

In Eq.\ \ref{eq4.5} we separated the probability density into a part depending
on the distances between the $z$-values ($l_i$ and $r_i$) and another part
containing the $x$-positions of motors.  Then we determined the attachment and
detachment rates (\ref{eq4.7},\ref{eq4.8}).  Now we describe the temporal
development of $P$ in terms of Master equations.  The detachment/attachment of
one specific site leads to destruction respectively creation of the following
states (see also Fig.\ \ref{fig8}):

\begin{displaymath}\begin{array}[t]{c|c}
\multicolumn{2}{l}{\mbox{Detachment}}\\ \\
\mbox{destroyed}&\mbox{created}\\
\hline\\
(\ldots,l_2,l_1 ; x ; r_1,r_2,\ldots)&\\
(\ldots,l_3,l_2 ; x_{l1} ; l_1,r_1,\ldots)&
(\ldots,l_3,l_2 ; x_{l1} ; l_1+r_1,r_2,\ldots)\\
(\ldots,l_1,r_1 ; x_{r1} ; r_2,r_3,\ldots)&
(\ldots,l_2,l_1+r_1 ; x_{r1} ; r_2,r_3,\ldots)\\
\vdots&\vdots\\
\end{array}
\qquad
\begin{array}[t]{c|c}
\multicolumn{2}{l}{\mbox{Attachment}}\\ \\
\mbox{destroyed}&\mbox{created}\\
\hline\\
&(\ldots,l_2,l_1 ; x^n ; r_1,r_2,\ldots)\\
(\ldots,l_3,l_2 ; x_{l1} ; l_1+r_1,r_2,\ldots)&(\ldots,l_3,l_2 ; x_{l1} ;
l_1,r_1,\ldots)\\
(\ldots,l_2,l_1+r_1 ; x_{r1} ; r_2,r_3,\ldots)&(\ldots,l_1,r_1 ; x_{r1} ;
r_2,r_3,\ldots)\\
\vdots&\vdots\\
\end{array}
\end{displaymath}

The transition rates are given by Eqns.\ (\ref{eq4.7}, \ref{eq4.8}).  The
Master equation for $P$ is
\begin{eqnarray}
\label{eq_master}
\lefteqn{\frac d{dt}P(\xi,t)_{\ldots,\lambda_2,\lambda_1;\rho_1,\rho_2,\ldots}
p(\ldots,\lambda_1;\rho_1,\ldots)= \int dx  \int dl_1 \int dr_1 \ldots}\nonumber\\
\Bigl[&&r_d(\ldots,l_2,l_1 ; x,t ; r_1,r_2,\ldots)\Bigl(-\delta(\xi-x)
\left(\cdots\delta(\lambda_1-l_1)\delta(\rho_1-r_1)\cdots\right)\nonumber\\
&&+P(x,t)_{\ldots,l_3,l_2;l_1,r_1\ldots}
(-\cdots\delta(\lambda_2-l_3)\delta(\lambda_1-l_2)
\delta(\rho_1-l_1)\cdots+\cdots\delta(\lambda_2-l_3)\delta(\lambda_1-l_2)
\delta(\rho_1-(l_1+r_1))\delta(\rho_2-r_2)\cdots)\nonumber\\
&&+P(x,t)_{\ldots,l_1,r_1;r_2,r_3\ldots}
(-\cdots\delta(\lambda_2-l_1)\delta(\lambda_1-r_1)
\delta(\rho_1-r_2)\cdots+\cdots\delta(\lambda_2-l_2)\delta(\lambda_1-(l_1+r_1))
\delta(\rho_1-r_2)\delta(\rho_2-r_3)\cdots)\nonumber\\ 
&&\vdots\nonumber\\
&&\Bigr)\nonumber\\ 
&&+r_a(\ldots,l_2,l_1 ; x,t ; r_1,r_2,\ldots)\Bigl(\delta(\xi-x)
\left(\cdots\delta(\lambda_1-l_1)\delta(\rho_1-r_1)\cdots\right)\nonumber\\
&&+P(x,t)_{\ldots,l_3,l_2;l_1+r_1,r_2\ldots}
(-\cdots\delta(\lambda_1-l_2)
\delta(\rho_1-(l_1+r_1))\delta(\rho_2-r_2)\cdots+\cdots\delta(\lambda_1-l_2)
\delta(\rho_1-l_1)\delta(\rho_2-r_1)\cdots)\nonumber\\
&&+P(x,t)_{\ldots,l_2,l_1+r_1;r_2,r_3\ldots}
(-\cdots\delta(\lambda_2-l_2)\delta(\lambda_1-(l_1+r_1))
\delta(\rho_1-r_2)\cdots+\cdots\delta(\lambda_2-l_1)\delta(\lambda_1-r_1) \delta(\rho_1-r_2)\cdots)\nonumber\\
&&\vdots\nonumber\\ &&\Bigr)\Bigr]
\end{eqnarray}
Rather than in the distribution itself we are interested in the expectation
value $\left< x (t) \right>_{\ldots,l_2,l_1 ; r_1,r_2,\ldots}$. Its equation of
motion follows directly form the Master equation.
\begin{eqnarray}
\label{eq_a2}
\lefteqn{\frac d{dt} \langle x(t)\rangle_{\ldots,\lambda_2,\lambda_1;\rho_1,\rho_2,\ldots}
p(\ldots,\lambda_1;\rho_1,\ldots)= \frac 1 {t_a} \int dl_1 \int dr_1 \ldots
p(\ldots,l_2,l_1 ; r_1,r_2,\ldots)}\nonumber\\
\Bigl[&-&\langle x (t) \rangle_{\ldots,l_2,l_1 ; r_1,r_2,\ldots}\delta(\xi-x)
\left(\cdots\delta(\lambda_1-l_1)\delta(\rho_1-r_1)\cdots\right)\nonumber\\
&+&\langle x (t) \rangle_{\ldots,l_3,l_2;l_1,r_1\ldots}
(-\cdots\delta(\lambda_2-l_3)\delta(\lambda_1-l_2)
\delta(\rho_1-l_1)\cdots+\cdots\delta(\lambda_2-l_3)\delta(\lambda_1-l_2)
\delta(\rho_1-(l_1+r_1))\delta(\rho_2-r_2)\cdots)\nonumber\\
&+&\langle x (t) \rangle_{\ldots,l_1,r_1;r_2,r_3\ldots}
(-\cdots\delta(\lambda_2-l_1)\delta(\lambda_1-r_1)
\delta(\rho_1-r_2)\cdots+\cdots\delta(\lambda_2-l_2)\delta(\lambda_1-(l_1+r_1))
\delta(\rho_1-r_2)\delta(\rho_2-r_3)\cdots)\nonumber\\ 
&&\vdots\nonumber\\
&+&\langle x^n (t) \rangle_{\ldots,l_2,l_1 ; r_1,r_2,\ldots}\delta(\xi-x)
\left(\cdots\delta(\lambda_1-l_1)\delta(\rho_1-r_1)\cdots\right)\nonumber\\
&+&\langle x (t) \rangle_{\ldots,l_3,l_2;l_1+r_1,r_2\ldots}
(-\cdots\delta(\lambda_1-l_2)
\delta(\rho_1-(l_1+r_1))\delta(\rho_2-r_2)\cdots+\cdots\delta(\lambda_1-l_2)
\delta(\rho_1-l_1)\delta(\rho_2-r_1)\cdots)\nonumber\\
&+&\langle x (t) \rangle_{\ldots,l_2,l_1+r_1;r_2,r_3\ldots}
(-\cdots\delta(\lambda_2-l_2)\delta(\lambda_1-(l_1+r_1))
\delta(\rho_1-r_2)\cdots+\cdots\delta(\lambda_2-l_1)\delta(\lambda_1-r_1) \delta(\rho_1-r_2)\cdots)\nonumber\\
&&\vdots\nonumber\\ &&\Bigr]
\end{eqnarray}
\beginmcols
This equation simplifies further if the distances $l_i$ and $r_i$ are
distributed according to their equilibrium distribution (\ref{eq4.6}),
which is certainly the case after the motors have been running for some time.  

Usually one is looking for the quasistationary solution with
\begin{displaymath}
\frac d{dt} \langle
x(t)\rangle_{\ldots,\lambda_2,\lambda_1;\rho_1,\rho_2,\ldots}=v\;.
\end{displaymath}
In the special case when $\langle x^n \rangle_{\ldots,l_2,l_1 ;
r_1,r_2,\ldots}$ does not depend on $l_i$ and $r_i$, the equation simplifies to
(\ref{eq4.13}).  Taking first-order correlations into account but
neglecting the higher ones leads to Eq.\ (\ref{eq4.25}).


\endmcols

\begin{thebibliography}{10}

\bibitem{alberts}
B. Alberts {\it et~al.}, {\em Molecular Biology of the cell}, 3 ed. (Garland
  Publ., New York, 1994).

\bibitem{howard97}
J. Howard, Nature {\bf 389},  561  (1997).

\bibitem{leibler93}
S. Leibler and D. Huse, J.~Cell.~Biol. {\bf 121},  1357  (1993).

\bibitem{hill38}
A. Hill, Proc. R. Soc. London Ser. B {\bf 126},  136  (1939).

\bibitem{huxley57}
A. Huxley, Prog. Biophys. Biophys. Chem. {\bf 7},  255  (1957).

\bibitem{huxley71}
A. Huxley and R. Simmons, Nature {\bf 233},  533  (1971).

\bibitem{toyoshima87}
Y. Toyoshima {\it et~al.}, Nature {\bf 328},  536  (1987).

\bibitem{howard89}
J. Howard, A. Hudsepth, and R. Vale, Nature {\bf 342},  154  (1989).

\bibitem{svoboda93}
K. Svoboda, C. Schmidt, B. Schnapp, and S. Block, Nature {\bf 365},  721
  (1993).

\bibitem{finer94}
J. Finer, R. Simmons, and J. Spudich, Nature {\bf 386},  113  (1994).

\bibitem{yanagida95}
T. Yanagida and A. Ishijima, Biophys.~J. {\bf 68},  312s  (1995).

\bibitem{ajdari92}
A. Ajdari and J. Prost, C.R. Acad.~Sci.~Paris~II {\bf 315},  1635  (1992).

\bibitem{peskin95}
C. Peskin and G. Oster, Biophys.~J. {\bf 68},  202s  (1995).

\bibitem{duke96}
T. Duke and S. Leibler, Biophys.~J. {\bf 71},  1235  (1996).

\bibitem{derenyi96}
I. Der{\'e}nyi and T. Vicsek, Proc.\ Natl.\ Acad.\ Sci.\ USA {\bf 93},  6775
  (1996).

\bibitem{prost95}
F. J{\"u}licher and J. Prost, Phys.~Rev.~Lett. {\bf 75},  2618  (1995).

\bibitem{juelicher97b}
F. J{\"u}licher, A. Ajdari, and J. Prost, Rev.~Mod.~Phys. {\bf 69},  1269
  (1997).

\bibitem{csahok97}
Z. Csah\'ok, F. Family, and T. Vicsek, Phys.~Rev.~E {\bf 55},  5179  (1997).

\bibitem{huxley94}
H. Huxley, A. Stewart, H. Sosa, and T. Irving, Biophys.~J. {\bf 67},  2411
  (1994).

\bibitem{wakabayashi94}
K. Wakabayashi {\it et~al.}, Biophys.~J. {\bf 67},  2422  (1994).

\bibitem{higuchi95}
H. Higuchi, T. Yanagida, and Y. Goldman, Biophys.~J. {\bf 69},  1000  (1995).

\bibitem{finer95}
J. Finer, A. Mehta, and J. Spudich, Biophys.~J. {\bf 68},  291s  (1995).

\bibitem{riveline98}
D. Riveline {\it et~al.}, European Biophys.~J.  (1998), in press.

\bibitem{leibler94}
S. Leibler, Nature {\bf 370},  412  (1994).

\bibitem{vilfan_preprint}
A. Vilfan, E. Frey, and F. Schwabl,   (1997), {preprint, cond-mat/9708023}.

\bibitem{note_applet}
A Java applet graphically visualising the simulation of our model is accessible
  on the internet address {\tt
  http://www.physik.tu-muenchen.de/\~{}avilfan/ecmm/}.

\bibitem{fenn24}
W. Fenn, J.~Physiol.\ (London) {\bf 184},  373  (1924).

\bibitem{juelicher97}
F. J{\"u}licher and J. Prost, Phys.~Rev.~Lett. {\bf 78},  4510  (1997).

\bibitem{ishijima96}
A. Ishijima {\it et~al.}, Biophys.~J. {\bf 70},  383  (1996).

\bibitem{mackintosh-kaes-janmey:95}
F. MacKintosh, J. K{\"a}s, and P. Janmey, Phys.~Rev.~Lett. {\bf 75},  4425
  (1995).

\bibitem{kroy-frey:96}
K. Kroy and E. Frey, Phys.~Rev.~Lett. {\bf 77},  306  (1996).

\bibitem{wilhelm-frey:96}
J. Wilhelm and E. Frey, Phys.~Rev.~Lett. {\bf 77},  2581  (1996).

\bibitem{frey-etal:condmat}
E. Frey, K. Kroy, J. Wilhelm, and E. Sackmann, Statistical Mechanics of
  Semiflexible Polymers: Theory and Experiments.

\bibitem{kojima94}
H. Kojima, A. Ishijima, and T. Yanagida, Proc.\ Natl.\ Acad.\ Sci.\ USA {\bf
  91},  12962  (1994).

\bibitem{riveline-etal:97}
D. Riveline, C.~H. Wiggins, R.~E. Goldstein, and A. Ott, Phys.~Rev.~E {\bf 56},
     (1997).

\bibitem{bagshaw93}
C. Bagshaw, {\em Muscle Contraction} (Chapman \& Hall, London, 1993).

\end{thebibliography}
\end{document}